\documentclass[12pt]{article} 

\usepackage{amsmath} 

\title{Can Fermi energy be determined by heating and/or cooling a copper wire? }
\author{A.C. Abhyankar\\
Department of Materials Engineering\\
Defence Institute of Advanced Technology\\
Pune 411025, India\\
and\\
Chetan Kotabage\\
Department of Physics\\
KLS Gogte Institute of Technology\\
Belgavi 590008, India\\
Email: cvkotabage@git.edu}

\begin{document}
\maketitle
\begin{center}
Abstract
\end{center}
 
Fermi energy, which is a theoretical concept, has been explored in an experiment for a copper wire \cite{1},\cite{3}.  In this article, scientific validity of this experiment is discussed. A correct method to determine Fermi energy is also discussed.

\section{Introduction}
The most simple model that describes conducting electrons in a metal considers no interaction between electrons. Introduction of quantum mechanical ideas into this model gives rise to a concept of Fermi energy of a metal. In metals, valence electrons are loosely bound to the nucleus. Due to this, valence electrons can easily get dissociated from the parent atom leaving behind a positive ion.  The Coulombic repulsion between large number of free electrons can be considered as balanced by attraction by the positive ions. The screening of positive charge occurs because of increased negative charge in its surroundings. Hence under such effects, free electrons can be assumed to be moving in a region of constant potential energy instead of a region of periodic Coulombic potential. Under such conditions, electrons can be treated as particles confined in a three dimensional infinite potential well.  The highest  energy level occupied by electrons when all lower energy levels are filled is known as Fermi energy.

The average number of Fermions at temperature $T$ in a state of energy $E$ is given by 
\begin{equation}
n_F=\frac{1}{e^{(E-E_F)/KT}+1}\;.
\end{equation} 
At T=0 K, as pointed out earlier, the electrons fill up the energy levels up the the Fermi energy $E_F$.  
\section{Fermi energy of the metal}

As discussed in \cite{2}, the Fermi energy, $E_F$, of metal is given as

\begin{equation}
E_F=\frac{\hbar^2}{2m}(3 \pi^2 n)^{2/3} \label{fermi-energy}\;,
\end{equation}
where $n$ is the density of electrons, $m$ is the mass of electron and $\hbar=h/2\pi$. 

If this model is applied to a metal, Fermi energy of a metal can be calculated by measuring density of electrons in a Hall effect experiment. 
 
\section{Fermi energy of copper}
The experimental method to determine Fermi energy is described in \cite{3}. In this method, the Fermi velocity of electron is expressed in terms of mean free path, $\lambda_F$, and relaxation time, $\tau$ as
\begin{equation}
v_F=\frac{\lambda_F}{\tau}\label{Fermi velocity}\;.
\end{equation}
Thus, the Fermi energy is expressed as
\begin{equation}
E_F=\frac{m\lambda_F^2}{2\tau^2}\;.
\end{equation}
 The relation between relaxation time and conductivity 
 \begin{equation}
 \tau=\frac{m\sigma}{ne^2} \label{collision time}
 \end{equation}
is utilized to get Fermi energy as
\begin{equation}
E_F=\Big(\frac{ne^2\lambda_F R \pi r^2}{L\sqrt{2 m}}\Big)^2\;,
\end{equation}
where conductivity, $\sigma=L/R\pi r^2$, with $L$ is the length of the copper wire, $r$ is the radius of wire and $R$ is the resistance of the wire is utilized. It appears that above formula has been rewritten as
\begin{equation}
E_F=\Big(\frac{ne^2\lambda_F T \pi r^2}{L\sqrt{2 m}}\Big)^2\times \Big[\frac{\Delta R}{\Delta T}\Big]^2\;, \label{Fermi-resistance}
\end{equation}
 to include temperature dependence of resistance of metals. With reference temperature, $T=318$ K, Fermi energy of various metals, including that of copper, have been reported in \cite{3}. 

\section{Drawbacks of the method}

The above method has the following drawbacks.

\subsection{Mean free path $\lambda_F$ and relaxation time}
The relaxation time is usually referred as the time taken by a free electron to decrease its velocity, $v_d$, by a factor of $1/e$ , when the electric field is turned off. The drift velocity of electron decreases as the electric field is switch off. In case of eq. \eqref{collision time}, relaxation time is defined when the metal wire is subjected to electrical potential. 

Apart from this, the relation between Fermi velocity and relaxation time in eq. \eqref{Fermi velocity} is inappropriate as it combines two different concepts of drift velocity and Fermi velocity. So, application of eq. \eqref{collision time} to calculate mean free path,$\lambda_F$, is incorrect. 

 Besides, $\lambda_F$  is calculated for Fermi velocity, which is obtained from theoretical value of Fermi energy of copper. This approach is also incorrect as the aim of the experiment is to find Fermi energy experimentally. 
\subsection{Fermi energy and density of electrons}
Since expression in eq. \eqref{Fermi-resistance} is obtained  from inappropriate application of relaxation time, the equation is invalid to get the Fermi energy.  As shown is eq. \eqref{fermi-energy}, measurement of density of electrons is enough to calculate Fermi energy of a metal.  Surprisingly, eq. \eqref{Fermi-resistance} uses theoretical value of density of electrons to estimate Fermi energy.  

Despite these errors, how are correct theoretical values for Fermi energy being reported in the experiment carried out in\cite{3}? The key is that mean free path, $\lambda_F$, is calculated from the theoretical value of Fermi energy. If eq. \eqref{Fermi-resistance} is rearranged as
\begin{equation}
E_F=\frac{v_F^2}{\tau^2}\Big(\frac{ne^2T \pi r^2}{L\sqrt{2 m}}\Big)^2\times \Big[\frac{\Delta R}{\Delta T}\Big]^2\;, \label{Fermi-resistance-1}
\end{equation} 
then the equation reduces to $E_F=mv_F^2/2$. Thus, the measurements that are carried out are irrelevant to get the final result because these measurements get cancelled in calculations.


\section{Conclusions}
The heating and/or cooling of a copper coil to study the dependence of resistance on temperature is a pointless exercise to determine Fermi energy. The correct way of finding Fermi energy is to measure density of electrons in Hall effect experiment. 

Another approximate way of finding density of electrons at room temperature is to utilize mobility, $\mu$, of electrons at room temperature. The current density and mobility of electrons can give the electron density as
\begin{equation}
n=\frac{IL}{e\mu V\pi r^2}\;,
\end{equation}
where $V$ is electric potential applied to the wire. By utilizing this density of electrons, Fermi energy at room temperature can be estimated.



%
%
%
%
\end{document}